\documentclass[%
preprint,
superscriptaddress,
 amsmath,amssymb,
 aps,
 prb,
]{revtex4-2}

\usepackage{graphicx}% Include figure files
\usepackage{dcolumn}% Align table columns on decimal point
\usepackage{bm}% bold math

\usepackage{color}
\usepackage{ulem}
\newcommand{\add}[1]{\textcolor{black}{#1}}
\newcommand{\del}[1]{}
\newcommand{\tsup}[1]{\ensuremath{^{#1}}}

\begin{document}

%\preprint{APS/123-QED}

%\title{Type-I and Type-II Saddle Points and a Topological Flat Band in a Bi-Pyrochlore Superconductor Cs$\textrm{Bi}_{2}$}% Force line breaks with \\
\begin{center}
{\large\bfseries
Type-I and Type-II Saddle Points and a Topological Flat Band in\\
a Bi-Pyrochlore Superconductor CsBi$_2$
\par}
\vspace{1.2em}

{\normalsize
Yusei Morita\tsup{1,*},
Yongkai Li\tsup{2,4,5,*},
Yu-Hao Wei\tsup{3,*},
Kosuke Nakayama\tsup{1,\dagger},
Zhiwei Wang\tsup{2,4,5,6,\dagger},
\add{Hua-Yu Li\tsup{3}},
Takemi Kato\tsup{7},
Seigo Souma\tsup{7,8},
Kiyohisa Tanaka\tsup{9,10},
\add{Kenichi Ozawa\tsup{11}},
\add{Jia-Xin Yin\tsup{12,13}},
Takashi Takahashi\tsup{1},
Min-Quan Kuang\tsup{3,\dagger},
Yugui Yao\tsup{2,4,5,6},
and Takafumi Sato\tsup{1,7,8,14,15,\dagger}
\par}
\vspace{1.0em}

{\normalsize\itshape
\tsup{1} Department of Physics, Graduate School of Science, Tohoku University, Sendai 980-8578, Japan\par
\tsup{2} Centre for Quantum Physics, Key Laboratory of Advanced Optoelectronic Quantum Architecture and Measurement (MOE), School of Physics, Beijing Institute of Technology, Beijing 100081, P. R. China\par
\tsup{3} Chongqing Key Laboratory of Micro \& Nano Structure Optoelectronics, and School of Physical Science and Technology, Southwest University, Chongqing 400715, P. R. China\par
\tsup{4} Beijing Key Lab of Nanophotonics and Ultrafine Optoelectronic Systems, Beijing Institute of Technology, Beijing 100081, P. R. China\par
\tsup{5} Material Science Center, Yangtze Delta Region Academy of Beijing Institute of Technology, Jiaxing 314011, P. R. China\par
\tsup{6} Beijing Institute of Technology, Zhuhai 519000, P. R. China\par
\tsup{7} Advanced Institute for Materials Research (WPI-AIMR), Tohoku University, Sendai 980-8577, Japan\par
\tsup{8} Center for Science and Innovation in Spintronics (CSIS), Tohoku University, Sendai 980-8577, Japan\par
\tsup{9} UVSOR Synchrotron Facility, Institute for Molecular Science, Okazaki 444-8585, Japan\par
\tsup{10} School of Physical Sciences, The Graduate University for Advanced Studies (SOKENDAI), Okazaki 444-8585, Japan\par
\tsup{11} Institute of Materials Structure Science, High Energy Accelerator Research Organization (KEK), Tsukuba 305-0801, Japan\par
\tsup{12} Department of Physics and Guangdong Basic Research Center of Excellence for Quantum Science, Southern University of Science and Technology, Shenzhen 518055, China\par
\tsup{13} Quantum Science Center of Guangdong-Hong Kong-Macao Greater Bay Area, Shenzhen, China\par
\tsup{14} International Center for Synchrotron Radiation Innovation Smart (SRIS), Tohoku University, Sendai 980-8577, Japan\par
\tsup{15} Mathematical Science Center for Co-creative Society (MathCCS), Tohoku University, Sendai 980-8578, Japan\par
\vspace{0.8em}

%\tsup{*} These authors contributed equally to this work.\par
%\tsup{\dagger} Corresponding authors: k.nakayama@arpes.phys.tohoku.ac.jp; zhiweiwang@bit.edu.cn; mqkuang@swu.edu.cn; t-sato@arpes.phys.tohoku.ac.jp
\par}
\end{center}

\date{\today}
\newpage
\begin{abstract}
The divergence of the electron density of states (DOS) plays an important role in enhancing many-body interactions and inducing various quantum phases in low-dimensional systems. However, such unique electronic structures remain experimentally elusive in three-dimensional (3D) systems, particularly those with strong spin-orbit coupling (SOC). Using angle-resolved photoemission spectroscopy and first-principles calculations for a Laves-phase superconductor CsBi$_2$, which features a Bi-pyrochlore 3D network with strong SOC, we identify two characteristic electronic structures with a large DOS. One is a dispersionless topological flat band \add{with $p$-orbital character,} locally formed around the U-K line, \add{which enhances DOS near the Fermi level}. The other involves type-I and type-II saddle points connected by a flat band, which cooperatively produce \add{an enhancement} in the DOS. Our findings suggest a novel mechanism for achieving a DOS \add{enhancement} and lay a foundation for exploring exotic phenomena driven by the interplay of multiple singularities with a large DOS, nontrivial topology, and strong SOC in 3D pyrochlores.
\end{abstract}
\maketitle
\vspace*{\fill}
\hrule height 0.4pt
\vspace{0.6em}
{\small
\noindent \tsup{*} These authors contributed equally to this work.\par
\noindent \tsup{\dagger} Corresponding authors: \par k.nakayama@arpes.phys.tohoku.ac.jp;\par  zhiweiwang@bit.edu.cn;\par
mqkuang@swu.edu.cn;\par t-sato@arpes.phys.tohoku.ac.jp\par
}
\clearpage  % ←本文は必ず次ページへ

%\section{Introduction}
Exploring materials and mechanisms that produce a divergent density of states (DOS) is a frontier in condensed-matter physics research. Achieving a divergent DOS becomes increasingly challenging as the dimensionality increases. For instance, in one-dimensional (1D) systems like nanotubes, the energy extremum of a simple parabolic band can cause a divergent DOS \cite{WilderNature1998,OdomNature1998}, but this no longer occurs in two-dimensional (2D) systems. A promising source of a large DOS in 2D and quasi-2D materials is a saddle point (SP), characterized by concave and convex energy dispersions along different directions in momentum space. The SP generates a van Hove singularity (VHS) with a logarithmic divergence in the DOS \cite{HovePR1953}, as in cuprates, pnictides, ruthenates, and kagome metals \cite{DessauPRL1993,GofronPRL1994,BorisenkoPRL2010,TamaiPRL2008,KangNP2022,HuNC2022}, potentially promoting various instabilities such as high-temperature superconductivity and unconventional density waves \cite{HePRX2021,FangPRB2015,PhanPRB2017,SunkoQM2019,BarberPRL2018,JiangNM2021,ZhouPRB2021,NakayamaPRB2021,LuoNC2023,TengNP2023}. Another key source of high DOS is a flat band \cite{MielkeJP1991,TasakiPRL1992,MielkeCMP1993,RegnaultNature2022,CalugaruNP2022}, which originates from interference of extended wave functions within special crystalline lattices, such as the kagome lattice and magic-angle twisted bilayer graphene \cite{CaoNature2018,CaoNature2018_2,UtamaNP2021,LisiNP2021,JobstNP2021,LinPRL2018,KangNM2020,LiuNC2020}. Flat bands in these systems have been a focus of intensive research because of their presumed links to novel correlated insulating, superconducting, magnetic, and strange-metal states \cite{SharpeScience2019,JiangNature2019,CaoPRL2020,ChoiNature2021,CaoScience2021,GaoPRX2023,YeNP2024}.

In contrast to 1D and 2D cases, three-dimensional (3D) platforms with a divergent DOS are quite scarce. A possible material class is the pyrochlore lattice, a 3D extension of the kagome lattice. Similar to the kagome lattice, the pyrochlore lattice is predicted to host SPs and flat bands, but with 3D characteristics \cite{BergmanPRB2008,GuoPRL2009}. While 3D SPs typically do not cause a DOS divergence \cite{HovePR1953}, 3D flat bands can, as recently reported in pyrochlores \cite{WakefieldNature2023,HuangNP2024,HuangAX2023}. However, the ideal flat-band character is present only in systems with relatively weak spin-orbit coupling (SOC), as strong SOC is predicted to increase the flat-band width. Therefore, realizing a divergent DOS in pyrochlores with strong SOC, which could lead to intriguing correlated topological and fractionalized properties \cite{PesinNP2010,WeeksPRB2012,MaciejkoNP2015,ZhouPRB2019}, remains as an important challenge.

In this Letter, we report a novel type of DOS \add{enhancement} in a 3D pyrochlore with strong SOC, CsBi$_2$, by conducting density functional theory (DFT) calculations and angle-resolved photoemission spectroscopy (ARPES) measurements \add{(see Supplemental Material \S1 for details of calculations and ARPES measurements)}. This \add{enhancement} originates from the coexistence of type-I and type-II SPs, defined as SPs at time-reversal-invariant momentum (TRIM) point and non-TRIM point, respectively \cite{MengPRB2015,YaoPRB2015}, and a topological flat band connecting these SPs. We also demonstrate an enhancement of the DOS near the Fermi energy ($E_F$) due to a flat band formed in specific momentum space. Our findings pave the way for exploring a large DOS and associated unconventional properties in 3D systems with strong SOC.

%\section{Results and Discussion}
$\textit{A}\textrm{Bi}_{2}$ ($\textit{A}$ = K, Rb, and Cs) \cite{PhilipPRB2023, GutowskaJPCC2023} is a C15-type Laves-phase material that crystallizes in a cubic structure ($Fd\bar{3}m$ space group; No. 227), with $\textit{A}$ and Bi atoms arranged in diamond and pyrochlore sublattices, respectively [Fig. 1(\add{a}); see Fig. 1(\add{b}) for the corresponding Brillouin zone (BZ)]. $\textit{A}\textrm{Bi}_{2}$ exhibits intriguing properties, including strong SOC of the Bi-pyrochlore sublattice, superconductivity ($\textit{T}_\mathrm{c}$ $\sim$ 4 K), and nonsaturating large magnetoresistance \cite{PhilipPRB2023, GutowskaJPCC2023, SunJPCM2016, PhilipPRB2021, LiPCCP2022}. \add{We first built the electronic tight-binding (TB) Hamilton for the 16c Wyckoff positions of space group Fd3m (No. 227) of CsBi$_2$ by the MagneticTB package \add{\cite{ZhangCPC2022}}. For the case of $s$ orbitals without SOC, the TB band dispersion presented in Fig. 1(c) shows an ideal two-fold degenerate flat band (labeled as bands 3 and 4) crossing the entire BZ, along with two dispersive bands (bands 1 and 2) that are symmetric around $-2$ eV. In addition to the SPs (black circles) located at the L point for the bands 1 and 2 (hereafter each SP is labeled by the location of the high-symmetry point with band No., e.g., L1 and L2 for these SPs), further SPs are identified at K1, U1, K2 and U2 (see Supplemental Material \S2 for the 3D band structures of these SPs). When SOC is included for $s$ orbitals, the corresponding TB band structure in Fig. 1(d) reveals that the originally flat band becomes dispersive, however, quasi-flat segments locally emerge along the U-K high-symmetry line. Moreover, the SPs identified in the SOC-free case are preserved, and additional SPs appear, as marked by black dots. Importantly, although the flat-band width is globally broadened by SOC, DOS remains nearly divergent at energies near the locally-formed flat bands and newly-emerged SPs around 2 and 4 eV.}

\add{To verify such large DOS in a real material, we performed} our band-structure calculations with SOC for CsBi$_2$. As confirmed by previous  band-structure calculations for $\textit{A}\textrm{Bi}_{2}$ and ARPES results for RbBi$_2$, there are multiple symmetry-enforced Dirac points (DPs) at the X point (indicated by red points) \cite{PhilipPRB2023,GutowskaJPCC2023,OhAX2024}, and two quadratic contact DPs at the $\Gamma$ point (orange points; see Supplemental Material \S\add{3} for details of the band topology). In addition, as expected from the presence of strong SOC, ideal flat bands are absent. However, calculated DOS in Fig. 1(f) shows notable enhancements characterized by a hump near $E_F$ and multiple spikes at higher binding energies ($E_{\textit{B}}$'s)\add{, similar to the high-DOS peaks in the TB model}.

The origins of the DOS enhancements are \add{twofold}. Firstly, examining the band dispersions reveals that nearly flat bands can exist in specific momentum cuts (e.g., along U-K) \add{like the TB model with SOC}, and one of such local flat bands (purple arrow) produces the hump near $E_F$. Secondly, analysis of the 3D band structures around high-symmetry points clarifies multiple SPs [black points in Fig. 1(e)], whose energies match some spikes in Fig. 1(f) (black arrows). These SPs are present not only at the TRIM point L (0.5, 0.5, 0.5), but also at non-TRIM points, such as W (0.5, 0.25, 0.75), U (0.625, 0.25, 0.625), and K (0.375, 0.375, 0.75). For instance, SPs at the L point characterized by opposite band velocities along two orthogonal momentum cuts, i.e., $\Gamma$-L and L-W, are identified for bands numbered as 2, 4, 12, 14, and 16 in Fig. 1(e). Similar SP characteristics are found at the W point for bands 4 and 6 [Fig. 1(g); see Supplemental Material \S4 for more details]. This unveils the coexistence of type-I SPs at TRIM points and type-II SPs at non-TRIM points \cite{MengPRB2015,YaoPRB2015}. An important consequence of the multiple SP formation is that the SPs at the same energy (U2 and K2, L4 and W4) generate one single logarithmic spike in the DOS [Fig. 1(f)], which echoes the prediction that two SPs corresponding to the same value of $E$($k$) might produce one single VHS in the DOS \cite{HovePR1953}. It is also noteworthy that the energetically degenerate SPs at the same branch are connected by a nearly dispersionless local flat band (U2-K2 and L4-W4), which would further contribute to the DOS enhancement.

\add{Importantly, our calculation of} $\mathbb{Z}_{2}$ \add{topological} indices (Table. 1) \add{for} the occupied bands \cite{FuPRB2007,FuPRL2007,FuPRL2011} indicate\add{s} that only band 8 is trivial and all others are nontrivial with nonzero $\mathbb{Z}_{2}$ indices, suggesting that the flattened bands along U2-K2, L4-W4, U10-K10, and U16-K16 paths are topologically nontrival. \add{This classification distinguishes them from trivial flat atomic bands derived from localized orbitals; instead, they are topological flat bands that cannot be described by localized orbitals (see Supplemental Material \S5 for further details)}. Therefore, despite the absence of ideal flat bands, locally-formed topological flat bands and unique SP degeneracy significantly enhance the DOS in CsBi$_2$. \add{Furthermore, while strong SOC is typically expected to broaden flat bands, our realistic band calculations with multi-orbitals clarify its constructive role. Notably, SOC induces orbital hybridizations that effectively produce sharper DOS peaks compared to the SOC-free case (see Supplemental Material \S6 for a comparison of band dispersions). The interplay of non-trivial topology and strong SOC is thus essential for the observed electronic singularities.}

To compare the calculated and experimental band structures in CsBi$_2$, we mapped out a 2D ARPES intensity. The map at $E_F$ [Fig. 2(a)] reveals a large Fermi surface at the BZ center and a small circular pocket at the BZ boundary. When increasing $E_{\textit{B}}$ to 0.3 eV [Fig. 2(b)], the former expands into a large hexagonal shape, while the latter disappears. These features originate from a highly-dispersive band and a shallow electron band, both of which cross $E_F$, as seen in Fig. 2(c) [cut along white dashed line in Fig. 2(a)]. We also conducted normal-emission measurements to clarify the three dimensionality of the band structure (see Supplemental Material \S\add{7}), and found that the data in Figs. 2(a)--2(c) reflect the electronic states near the $\textit{k}_{z} = \pi/3$ plane. As shown by the red curves in Fig. 2(c) and the calculated Fermi surface in Fig. 2(d), the highly-dispersive band and the shallow pocket are qualitatively reproduced by the DFT calculations for $\textit{k}_{z}$ = $\pi$/3 [see also Supplemental Material \S\add{8} for additional data suggesting the DP formation at the X point].

Next, to experimentally identify the near-$E_F$ flat band, one of key predictions from the DFT calculations, we measured the band dispersion along the U-K cut at $\textit{k}_{z}$ = $\pi$/6 [cut 2 in Fig. 2(e)]. Figure 2(f) shows several less-dispersive bands in this momentum region, which reasonably agree with the DFT calculations [compare orange circles and red curves in Fig. 2(g)\add{; see Supplemental Material \S9 for the numerical fitting to extract the peak position].} \add{We note} some quantitative differences such as a flattening of the band at $E_B$ $\sim$ 1.0 eV\add{, which is an extrinsic effect arising from angle-integrated signals generated by high DOS (see Supplemental Material \S10)}. \add{Importantly, the} band near $E_F$ corresponds to the predicted flat band along U16-K16, and produces a near-$E_F$ peak in the local DOS obtained by integrating energy distribution curves (EDCs) between U and K [Fig. 2(h)], consistent with the predicted DOS enhancement \add{and potentially responsible for the strong-coupling superconductivity \cite{SongPRB2025} (see also Supplemental Material \S11)}.

We next focus on another key prediction, the presence of type-I and type-II SPs at nearly the same $E_B$ and their connection by a flat band, whose characteristics would be most clearly seen along L4-W4 [see schematics in Figs. 3(a) and 3(b)]. Figures 3(c1)--3(c3) and 3(d1)--3(d3) show a comparison of experimental and calculated band dispersions along $\sim\!\Gamma$-L and L-K/U. Along $\sim\!\Gamma$-L, the experiment reveals relatively shallow electron pockets within 0.5 eV of $E_F$ and deeper holelike dispersions centered at the L point [Fig. 3(c1)], which overall agree with the calculations [Fig. 3(c2)]. Notably, band 4, located at $E_B$ = 3.2--3.4 eV, exhibits a holelike dispersion topped at the L point [black dashed line in Fig. 3(c1)], as confirmed by tracing the peak position in the magnified view of EDCs at $E_B$ = 3.0--3.5 eV [Fig. 3(c3)]. In contrast, along the L-K/U cut, this band has the bottom of an electronlike dispersion at the L point [Figs. 3(d1) and 3(d3)\add{; see Supplemental Material \S13 for a magnified view near $E_F$}]. The observed opposite band curvature with respect to the L point confirms the formation of a type-I SP at the TRIM point, in agreement with the calculations [Figs. 3(c2) and 3(d2)]. Furthermore, the existence of a type-II SP at the W point (non-TRIM) is supported by a comparison of Figs. 3(e1)--3(e3) and 3(f1)--3(f3), where band 4 shows the holelike and electronlike dispersion centered at W along $\sim\!\Gamma$-W and W-U, respectively. Remarkably, these two SPs are located at approximately the same $E_B$ of $\sim$3.25 eV [see black circles in Figs. 3(c3)--3(e3)] and connected by a \add{quasi-flat} band along the L-W path [Fig. 3(g1)--3(g3)], resulting in a clear peak in the momentum-integrated EDC in Fig. 3(h), consistent with the \add{enhancement} in the calculated DOS [Fig. 3(i)]. \add{Intriguingly, our analysis also identifies another SP at the U point (Supplemental Material \S14). According to the recent theoretical classification of singularities in 3D materials \cite{noncriticalSP}, the SP at the U point in pyrochlores is a noncritical SP, which represents a new class of electronic singularities.}

We now discuss the implications of the present results. Firstly, \add{our observation expands the scope of topological flat-band materials. Unlike the $d$-orbital flat bands reported in 3D systems such as CaNi$_2$ and CeRu$_2$ \cite{WakefieldNature2023,HuangAX2023}, the flat bands in CsBi$_2$ exhibit dominant $p$-orbital character [Fig. 2(c)]. This finding is particularly timely, as $p$-orbital flat bands have been recently predicted theoretically to host unique correlated phases \cite{ZhenPRB2024} but have remained experimentally elusive in 3D kagome/pyrochlore networks until now.} Secondly, \add{the identification of type-II SP is noteworthy because it has been theoretically linked to unconventional superconductivity, especially topological odd-parity superconductivity \cite{MengPRB2015,YaoPRB2015}. Therefore, it is an intriguing future work to explore pyrochlores hosting type-II SPs near $E_F$.} \add{Finally,} \add{the coexistence of} type-I and type-II SPs connected by a flat band \add{represents a new mechanism for DOS enhancement in correlated 3D systems with strong SOC}. These bands are at nearly the same energy and collectively lead to \add{an enhanced} DOS. This finding is unexpected from general hypotheses that (i) 3D SP-band systems do not exhibit such \add{an enhanced} DOS and (ii) strong SOC diminishes the ideal flat band character to broaden the singularity of the DOS in pyrochlores. \add{Since the essential features are well reproduced in the TB model [Fig. 1(d)], multiple singularities and resonant DOS enhancement can be commonly found in other pyrochlores.}

%\section{Conclusion}
In conclusion, we presented DFT and ARPES investigations of a Laves phase superconductor Cs$\textrm{Bi}_{2}$. The results demonstrated several unique band structures: (i) an incipient flat band close to $E_F$, (ii) multiple SPs with both type-I and type-II characteristics, and (iii) a topological flat band bridging two SPs in the same branch. \add{Multiple} SPs connected by a flat band lead to a DOS \add{enhancement}. These findings contribute to our deeper understanding of unique physical properties of CsBi$_2$, and suggests a new mechanism to realize a DOS \add{enhancement} and associated instabilities in correlated 3D systems with strong SOC.

\begin{acknowledgments}
We thank Miho Kitamura, Koji Horiba, and Hiroshi Kumigashira for their assistance in the ARPES experiments at Photon Factory. This work was supported by JST-CREST (No. JPMJCR18T1), Grant-in-Aid for Scientific Research (JSPS KAKENHI Grant Numbers JP21H04435, JP23KJ0099, and JP23K25812), UVSOR (Proposal numbers: 23IMS6835, 23IMS6846), and KEK-PF (Proposal numbers: 2024S2-001). The work at Beijing Institute of Technology was supported by the National Key R\&D Program of China (Grant Nos. 2020YFA0308800, 2022YFA1403400), \add{the Beijing National Laboratory for Condensed Matter Physics (Grant No. 2023BNLCMPKF007)}, and the Beijing Natural Science Foundation (Grant No. Z210006). \add{The work at Southwest University was supported by the Natural Science Foundation of Chongqing (Grant No. CSTB2024NSCQ-MSX0080) and the National Natural Science Foundation of China (NSFC, Grant No. 11704315).} Y.M. acknowledges support from GP-MS at Tohoku University \add{and JSPS}. T.K. acknowledges support from GP-Spin at Tohoku University and JST-SPRING (No. JPMJSP2114). Z.W. thanks the Analysis \& Testing Center at BIT for assistance in facility support. M.Q. Kuang acknowledges support from the high performance-computing platform located in School of Physical Science and Technology of Southwest University.
\end{acknowledgments}

\newpage
\bibliographystyle{prsty}

\begin{thebibliography}{50}

\bibitem{WilderNature1998} J. W. G. Wilder, L. C. Venema, A. G. Rinzler, R. E. Smalley, and C. Dekker, Nature \textbf{391}, 59-62 (1998).
\bibitem{OdomNature1998} T. W. Odom, J.-L. Huang, P. Kim, and C. M. Lieber, Nature \textbf{391}, 62-64 (1998).

\bibitem{HovePR1953} L. V. Hove, Phys. Rev. \textbf{89}, 1189 (1953).

\bibitem{DessauPRL1993} D. S. Dessau, Z.-X. Shen, D. M. King, D. S. Marshall, L. W. Lombardo, P. H. Dickinson, A. G. Loeser, J. DiCarlo, C.-H. Park, A. Kapitulnik, and W. E. Spicer, Phys. Rev. Lett. \textbf{71}, 2781 (1993).
\bibitem{GofronPRL1994} K. Gofron, J. C. Campuzano, A. A. Abrikosov, M. Lindroos, A. Bansil, H. Ding, D. Koelling, and B. Dabrowski, Phys. Rev. Lett. \textbf{73}, 3302 (1994).
\bibitem{BorisenkoPRL2010} S. V. Borisenko, V. B. Zabolotnyy, D. V. Evtushinsky, T. K. Kim, I. V. Morozov, A. N. Yaresko, A. A. Kordyuk, G. Behr, A. Vasiliev, R. Follath, and B. B\"uchner, Phys. Rev. Lett. \textbf{105}, 067002 (2010).
\bibitem{TamaiPRL2008} A. Tamai, M. P. Allan, J. F. Mercure, W. Meevasana, R. Dunkel, D. H. Lu, R. S. Perry, A. P. Mackenzie, D. J. Singh, Z.-X. Shen, and F. Baumberger, Phys. Rev. Lett. \textbf{101}, 026407 (2008).
\bibitem{KangNP2022} M. Kang, S. Fang, J.-K. Kim, B. R. Ortiz, S. H. Ryu, J. Kim, J. Yoo, G. Sangiovanni, D. D. Sante, B.-G. Park, C. Jozwiak, A. Bostwick, E. Rotenberg, E. Kaxiras, S. D. Wilson, J.-H. Park, and R. Comin, Nat. Phys. \textbf{18}, 301-308 (2022).
\bibitem{HuNC2022} Y. Hu, X. Wu, B.n R. Ortiz, S. Ju, X. Han, J. Ma, N. C. Plumb, M. Radovic, R. Thomale, S. D. Wilson, A. P. Schnyder, and M. Shi, Nat. Commun. \textbf{13}, 2220 (2022).

\bibitem{HePRX2021} Y. He, S.-D. Chen, Z.-X. Li, D. Zhao, D. Song, Y. Yoshida, H. Eisaki, T. Wu, X.-H. Chen, D.-H. Lu, C. Meingast, T. P. Devereaux, R. J. Birgeneau, M. Hashimoto, D.-H. Lee, and Z.-X. Shen, Phys, Rev. X \textbf{11}, 031068 (2021). 
\bibitem{FangPRB2015} D. Fang, X. Shi, Z. Du, P. Richard, H. Yang, X. X. Wu, P. Zhang, T. Qian, X. Ding, Z. Wang, T. K. Kim, M. Hoesch, A. Wang, X. Chen, J. Hu, H. Ding, and H.-H. Wen, Phys. Rev. B \textbf{92}, 144513 (2015).
\bibitem{PhanPRB2017} G. N. Phan, K. Nakayama, K. Sugawara, T. Sato, T. Urata, Y. Tanabe, K. Tanigaki, F. Nabeshima, Y. Imai, A. Maeda, and T. Takahashi, Phys. Rev. B \textbf{95}, 224507 (2017).
\bibitem{SunkoQM2019} V. Sunko, E. A. Morales, I. Markovi\'c, M. E. Barber, D. Milosavljevi\'c, F. Mazzola, D. A. Sokolov, N. Kikugawa, C. Cacho, P. Dudin, H. Rosner, C. W. Hicks, P. D. C. King, and A. P. Mackenzie, npj Quantum Mater. \textbf{4}, 46 (2019).
\bibitem{BarberPRL2018} M. E. Barber, A. S. Gibbs, Y. Maeno, A. P. Mackenzie, and C. W. Hicks, Phys.Rev. Lett. \textbf{120}, 076602 (2018).
\bibitem{JiangNM2021} Y.-X. Jiang, J.-X. Yin, M. M. Denner, N. Shumiya, B. R. Ortiz, G. Xu, Z. Guguchia, J. He, M. S. Hossain, X. Liu, J. Ruff, L. Kautzsch, S. S. Zhang, G. Chang, I. Belopolski, Q. Zhang, T. A. Cochran, D. Multer, M. Litskevich, Z.-J. Cheng, X. P. Yang, Z. Wang, R. Thomale, T. Neupert, S. D. Wilson, and M. Z. Hasan, Nat. Mater. \textbf{20}, 1353-1357 (2021).
\bibitem{ZhouPRB2021} X. Zhou, Y. Li, X. Fan, J. Hao, Y. Dai, Z. Wang, Y. Yao, and H.-H. Wen, Phys. Rev. B \textbf{104}, L041101 (2021).
\bibitem{NakayamaPRB2021} K. Nakayama, Y. Li, T. Kato, M. Liu, Z. Wang, T. Takahashi, Y. Yao, and T. Sato, Phys. Rev. B \textbf{104}, L161112 (2021).
\bibitem{LuoNC2023} Y. Luo, Y. Han, J. Liu, H. Chen, Z. Huang, L. Huai, H. Li, B. Wang, J. Shen, S. Ding, Z. Li, S. Peng, Z. Wei, Y. Miao, X. Sun, Z. Ou, Z. Xiang, M. Hashimoto, D. Lu, Y. Yao, H. Yang, X. Chen, H.-J. Gao, Z. Qiao, Z. Wang, and J. He, Nat. Commun. \textbf{14}, 3819 (2023).
\bibitem{TengNP2023} X. Teng, J. S. Oh, H. Tan, L. Chen, J. Huang, B. Gao, J.-X. Yin, J.-H. Chu, M. Hashimoto, D. Lu, C. Jozwiak, A. Bostwick, E. Rotenberg, G. E. Granroth, B. Yan, R. J. Birgeneau, P. Dai, and M. Yi, Nat. Phys. \textbf{19}, 814-822 (2023).

\bibitem{MielkeJP1991} A. Mielke, J. Phys. A \textbf{24}, L73-L77 (1991); J. Phys. A \textbf{24}, 3311-3321 (1991); J. Phys. A \textbf{25}, 4335-4345 (1992).
\bibitem{TasakiPRL1992} H. Tasaki, Phys. Rev. Lett. \textbf{69}, 1608-1611 (1992).
\bibitem{MielkeCMP1993}  A. Mielke and H. Tasaki, Commun. Math. Phys. \textbf{158}, 341-371 (1993).
\bibitem{RegnaultNature2022} N. Regnault, Y. Xu, M.-R. Li, D.-S. Ma, M. Jovanovic, A. Yazdani, S. S. P. Parkin, C. Felser, L. M. Schoop, N. P. Ong, R. J. Cava, L. Elcoro, Z.-D. Song, and B. A. Bernevig, Nature \textbf{603}, 824-828 (2022).
\bibitem{CalugaruNP2022} D. C\v{a}lug\v{a}ru, A. Chew, L. Elcoro, Y. Xu, N. Regnault, Z.-D. Song, and B. A. Bernevig, Nat. Phys. \textbf{18}, 185-189 (2022).

\bibitem{CaoNature2018} Y. Cao, V. Fatemi, A. Demir, S. Fang, S. L. Tomarken, J. Y. Luo, J. D. Sanchez-Yamagishi, K. Watanabe, T. Taniguchi, E. Kaxiras, R.C. Ashoori, P. Jarillo-Herrero, Nature \textbf{556}, 80-84 (2018).
\bibitem{CaoNature2018_2} Y. Cao, V. Fatemi, S. Fang, K. Watanabe, T. Taniguchi, E. Kaxiras, and P. Jarillo-Herrero, Nature \textbf{556}, 43-50 (2018).
\bibitem{UtamaNP2021} M. I. B. Utama, R. J. Koch, K. Lee, N. Leconte, H. Li, S. Zhao, L. Jiang, J. Zhu, K. Watanabe, T. Taniguchi, P. D. Ashby, A. Weber-Bargioni, A. Zettl, C. Jozwiak, J. Jung, E. Rotenberg, A. Bostwick, and F. Wang, Nat. Phys. \textbf{17}, 184-188 (2021).
\bibitem{LisiNP2021} S. Lisi, X. Lu, T. Benschop, T. A. de Jong, P. Stepanov, J. R. Duran, F. Margot, I. Cucchi, E. Cappelli, A. Hunter, A. Tamai, V. Kandyba, A. Giampietri, A. Barinov, J. Jobst, V. Stalman, M. Leeuwenhoek, K. Watanabe, T. Taniguchi, L. Rademaker, S. J. van der Molen, M. P. Allan, D. K. Efetov, and F. Baumberger, Nat. Phys. \textbf{17}, 189-193 (2021).
\bibitem{JobstNP2021} J. Jobst, V. Stalman, M. Leeuwenhoek, K. Watanabe, T. Taniguchi, L. Rademaker, S. J. van der Molen, M. P. Allan, D. K. Efetov, and F. Baumberger, Nat. Phys. \textbf{17}, 189-193 (2021).
\bibitem{LinPRL2018} Z. Lin, J.-H. Choi, Q. Zhang, W. Qin, S. Yi, P. Wang, L. Li, Y. Wang, H. Zhang, Z. Sun, L. Wei, S. Zhang, T. Guo, Q. Lu, J.-H. Cho, C. Zeng, and Z. Zhang, Phys. Rev. Lett. \textbf{121}, 096401 (2018).
\bibitem{KangNM2020} M. Kang, L. Ye, S. Fang, J.-S. You, A. Levitan, M. Han, J. I. Facio, C. Jozwiak, A. Bostwick, E. Rotenberg, M. K. Chan, R. D. McDonald, D. Graf, K. Kaznatcheev, E. Vescovo, D. C. Bell, E. Kaxiras, J. van den Brink, M. Richter, M. P. Ghimire, J. G. Checkelsky, and R. Comin, Nat. Mater. \textbf{19}, 163-169 (2020).
\bibitem{LiuNC2020} Z. Liu, M. Li, Qi Wang, G. Wang, C. Wen, K. Jiang, X. Lu, S. Yan, Y. Huang, D. Shen, J.-X. Yin, Z. Wang, Z. Yin, H. Lei, and S. Wang, Nat. Commun. \textbf{11}, 4002 (2020).

\bibitem{SharpeScience2019} A. L. Sharpe, E. J. Fox, A. W. Barnard, J. Finney, K. Watanabe, T. Taniguchi, M. A. Kastner, and D. Goldhaber-Gordon, Science \textbf{365}, 605 (2019).
\bibitem{JiangNature2019} Y. Jiang, X. Lai, K. Watanabe, T. Taniguchi, K. Haule, J. Mao, and E. Y. Andrei, Nature \textbf{573}, 91 (2019).
\bibitem{CaoPRL2020} Y. Cao, D. Chowdhury, D. Rodan-Legrain, O. Rubies-Bigorda, K. Watanabe, T. Taniguchi, T. Senthil, and P. Jarillo-Herrero, Phys. Rev. Lett. \textbf{124}, 076801 (2020).
\bibitem{ChoiNature2021} Y. Choi, H. Kim, Y. Peng, A. Thomson, C. Lewandowski, R. Polski, Y. Zhang, H. S. Arora, K. Watanabe, T. Taniguchi, J. Alicea, and S. Nadj-Perge, Nature \textbf{589}, 536 (2021).
\bibitem{CaoScience2021} Y. Cao, D. Rodan-Legrain, J. M. Park, N. F. Q. Yuan, K. Watanabe, T. Taniguchi, R. M. Fernandes, L. Fu, and P. Jarillo-Herrero, Science \textbf{372}, 264 (2021).
\bibitem{GaoPRX2023} S. Gao, S. Zhang, C. Wang, S. Yan, X. Han, X. Ji, W. Tao, J. Liu, T. Wang, S. Yuan, G. Qu, Z. Chen, Y. Zhang, J. Huang, M. Pan, S. Peng, Y. Hu, H. Li, Y. Huang, H. Zhou, S. Meng, L. Yang, Z. Wang, Y. Yao, Z. Chen, M. Shi, H. Ding, H. Yang, K. Jiang, Y. Li, H. Lei, Y. Shi, H. Weng, and T. Qian,  Phys. Rev. X \textbf{13}, 041049 (2023).
\bibitem{YeNP2024} L. Ye, S. Fang, M. Kang, J. Kaufmann, Y. Lee, C. John, P. M. Neves, S. Y. Frank Zhao, J. Denlinger, C. Jozwiak, A. Bostwick, E. Rotenberg, E. Kaxiras, D. C. Bell, O. Janson, R. Comin, and J. G. Checkelsky, Nat. Phys. \textbf{20}, 610-614 (2024).

\bibitem{MengPRB2015} Z. Y. Meng, F. Yang, K.-S. Chen, H. Yao, and H.-Y. Kee, Phys. Rev. B \textbf{91}, 184509 (2015).
\bibitem{YaoPRB2015} H. Yao and F. Yang, Phys. Rev. B \textbf{92}, 035132 (2015).

\bibitem{BergmanPRB2008} D. L. Bergman, C. Wu, and L. Balents, Phys. Rev. B \textbf{78}, 125104 (2008).
\bibitem{GuoPRL2009} H.-M. Guo and M. Franz, Phys. Rev. Lett. \textbf{103}, 206805 (2009).

\bibitem{WakefieldNature2023} J. P. Wakefield, M. Kang, P. M. Neves, D. Oh, S. Fang, R. McTigue, S. Y. Frank Zhao, T. N. Lamichhane, A. Chen, S. Lee, S. Park, J.-H. Park, C. Jozwiak, A. Bostwick, E. Rotenberg, A. Rajapitamahuni, E. Vescovo, J. L. McChesney, D. Graf, J. C. Palmstrom, T. Suzuki, M. Li, R. Comin, and J. G. Checkelsky, Nature \textbf{623}, 301 (2023).
\bibitem{HuangNP2024} J. Huang, L. Chen, Y. Huang, C. Setty, B. Gao, Y. Shi, Z. Liu, Y. Zhang, T. Yilmaz, E. Vescovo, M. Hashimoto, D. Lu, B. I. Yakobson, P. Dai, J.-H. Chu, Q. Si, and M. Yi, Nat. Phys. \textbf{20}, 603 (2024).
\bibitem{HuangAX2023} J. Huang, C. Setty, L. Deng, J.-Y. You, H. Liu, S. Shao, J. S. Oh, Y. Guo, Y. Zhang, Z. Yue, J.-X. Yin, M. Hashimoto, D. Lu, S. Gorovikov, P. Dai, J. D. Denlinger, M. Z. Hasan, Y.-P. Feng, R. J. Birgeneau, Y. Shi, C.-W. Chu, G. Chang, Q. Si, and M. Yi, arXiv:2304.09066.

\bibitem{PesinNP2010} D. Pesin and L. Balents, Nat. Phys. \textbf{6}, 376-381 (2010).
\bibitem{WeeksPRB2012} C. Weeks and M. Franz, Phys. Rev. B \textbf{85}, 041104(R) (2012).
\bibitem{MaciejkoNP2015} J. Maciejko and G. A. Fiete, Nat. Phys. \textbf{11}, 385-388 (2015).
\bibitem{ZhouPRB2019} Y. Zhou, K.-H. Jin, H. Huang, Z. Wang, and F. Liu, Phys. Rev. B \textbf{99}, 201105(R) (2019).

\bibitem{PhilipPRB2023} S. S. Philip, S. Liu, J. C. Y. Teo, P. V. Balachandran, and D. Louca, Phys. Rev. B \textbf{107}, 035143 (2023).
\bibitem{GutowskaJPCC2023} S. Gutowska, B. Wiendlocha, T. Klimczuk, and M. Winiarski, J. Phys. Chem. C \textbf{127}, 14402 (2023).
\bibitem{SunJPCM2016} S. Sun, K. Liu, and H. Lei, J. Phys.: Condens. Matter \textbf{28}, 085701 (2016).
\bibitem{PhilipPRB2021} S. S. Philip, J. Yang, D. Louca, P. F. S. Rosa, J. D. Thompson, and K. L. Page, Phys. Rev. B \textbf{104}, 104503 (2021).
\bibitem{LiPCCP2022} H. Li, M. Ikeda, A. Suzuki, T. Taguchi, Y. Zhang, H. Goto, R. Eguchi, Y.-F. Liao, H. Ishii, and Y. Kubozono, Phys. Chem. Chem. Phys. \textbf{24}, 7185-7194 (2022).
\add{\bibitem{ZhangCPC2022} Z. Zhang, Z.-M. Yu, G.-B. Liu, and Y. Yao, Comput. Phys. Commun. \textbf{270}, 108153 (2022).}
\bibitem{OhAX2024}  D. Oh, J. Kang, Y. Qian, S. Fang, M. Kang, C. Jozwiak, A. Bostwick, E. Rotenberg, J. G. Checkelsky, L. Fu, T. Klimczuk, M. J. Winiarski, B.-J. Yang, and R. Comin, arXiv:2402.04509.

\bibitem{FuPRB2007} L. Fu and C. L. Kane, Phys. Rev. B \textbf{76}, 045302 (2007).
\bibitem{FuPRL2007} L. Fu, C. L. Kane, and E. J. Mele, Phys. Rev. Lett. \textbf{98}, 106803 (2007).
\bibitem{FuPRL2011} L. Fu, Phys. Rev. Lett. \textbf{106}, 106802 (2011).
\add{\bibitem{SongPRB2025} W. Song, G. Liu, H. Deng, T. Yang, Y. Li, X.-Y. Yan, R. Liao, Q. Wang, J. Xu, C. Yan, Y. Zhao, H. Qin, D. Wang, W. Jing, D. Shen, K. Nakayama, T. Sato, C. Setty, D. Wu, B. Song, T. Ying, Z. Tian, A. Sakai, S. Nakatsuji, H. Kumar, C. A. Kuntscher, Z. Wang, Q.-K. Xue, and J.-X. Yin, Phys. Rev. B \textbf{112}, 245131 (2025).}
\add{\bibitem{noncriticalSP}H.-Y. Li, H. Tan, H.-Y. Zhu, H.-K. Yuan, and M.-Q. Kuang, arXiv:2604.07806.}
\add{\bibitem{ZhenPRB2024} Y. Zheng, L. Li, N. Luo, L.-M. Tang, Y. Feng, K.-Q. Chen, Z. Zhang, and J. Zeng, Phys. Rev. B. \textbf{109}, L180504 (2024).}

\bibitem{MommaJAC2011} K. Momma and F. Izumi, J. Appl. Cryst. \textbf{44}, 1272-1276 (2011).

\end{thebibliography}

\newpage

\begin{figure*}[htbp]
\includegraphics[width=165mm]{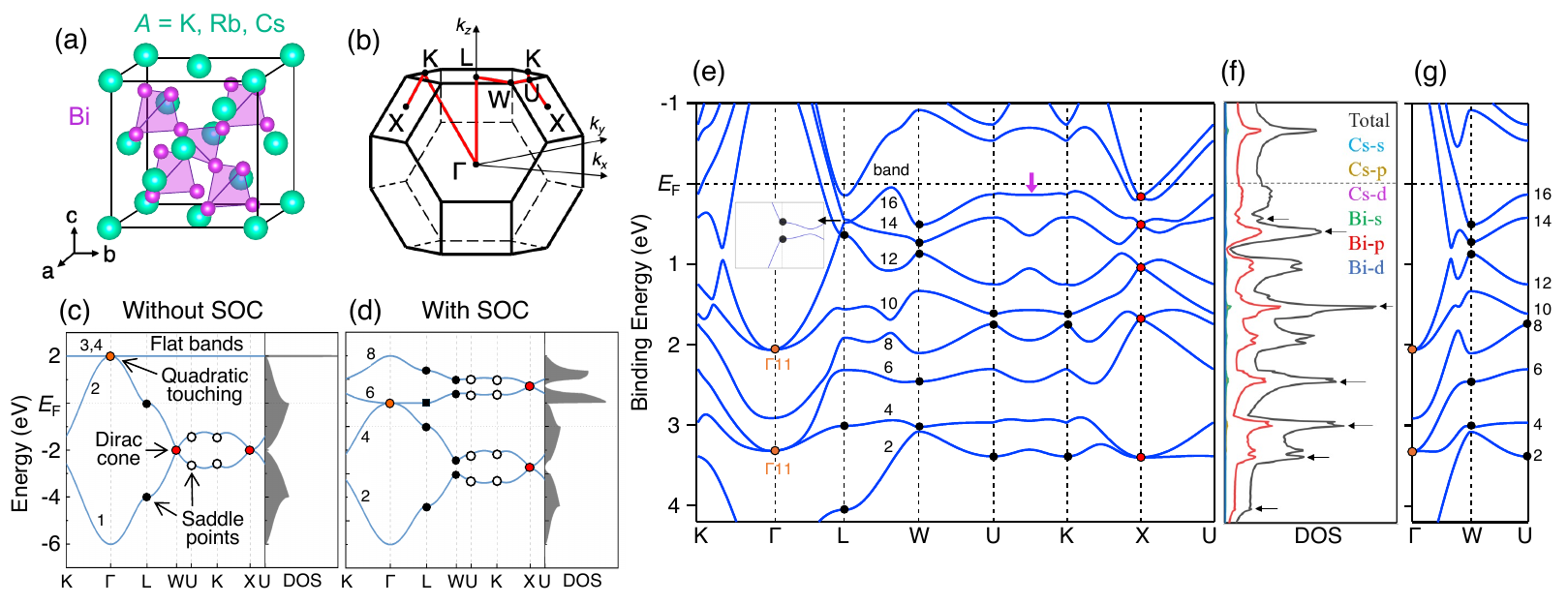}
 \caption{\label{Fig1}
\add{(a)} Crystal structure of cubic Cs$\textrm{Bi}_{2}$. Bi-pyrochlore network is highlighted by purple tetrahedrons. The graphic was drawn by VESTA \cite{MommaJAC2011}. \add{(b)} BZ of Cs$\textrm{Bi}_{2}$. \add{(c) and (d) Band structures of pyrochlore lattice (left) obtained by the TB model without and with SOC, respectively, together with the corresponding DOS (right) \cite{noncriticalSP}. The momentum paths are indicated by red lines in (b).} (e) Calculated band structure with SOC for Cs$\textrm{Bi}_{2}$. Occupied bands are sequentially numbered. Red, orange, and black dots indicate the positions of DP, quadratic touching, and SP, respectively. Inset shows an enlarged view for the two SPs at L near $E_F$. (f) Calculated orbital-resolved DOS. Black arrows highlight spikes due to SPs. (g) Same as (e), but along different momentum paths for visualizing type-II SPs (black dots) at W.}
 \end{figure*}
 
 \newpage
 
 %%%
\begin{table}[t]
\caption{Parity of the wave function at TRIM points, and the $\mathbb{Z}_{2}$-type topological invariant of CsBi$_2$. The coordinates of eight TRIM points are $\Gamma$ (0, 0, 0), X$_{1}$(0.5, 0.5, 0), X$_{2}$(0.5, 0, 0.5), X$_{3}$(0, 0.5, 0.5), L(0.5, 0.5, 0.5), F$_{1}$(0.5, 0, 0), F$_{2}$(0, 0.5, 0), and F$_{3}$(0, 0, 0.5). The + ($-$) denotes the positive (negative) parity eigenvalue.}
\vspace{0.1cm}
\renewcommand\arraystretch{1.3}
\setlength{\tabcolsep}{1.8mm}{
\begin{tabular}{c c c c c c c c c c c c}
\hline
\hline 
  & \multicolumn{4}{c}{Parity} &         & \multicolumn{4}{c}{Product of parity} &   \tabularnewline
 Band index & $\Gamma$ & 3X  & L & 3F &      &$\Gamma$ & 3X & L & 3F &   $\mathbb{Z}_{2} (\nu_{0},\nu_{1}\nu_{2}\nu_{3})$ \tabularnewline 
\hline 
2 & $-$ & +  & + & $-$ &   & $-$ & + & + & $-$ & (0,111) \tabularnewline
4 & $-$ & $-$  & + & $-$ &   & + & $-$ & + & + & (1,000) \tabularnewline
6 & $-$ & +  & + & $-$ &   & $-$ & $-$ & + & $-$ & (1,111) \tabularnewline
8 & $-$ & $-$  & $-$ & + &   & + & + & $-$ & $-$ & (0,000) \tabularnewline
10 & $-$ & +  & + & $-$ &   & $-$ & + & $-$ & + & (0,111) \tabularnewline
12 & $-$ & $-$  & + & $-$ &   & + & $-$ & $-$ & $-$ & (1,000) \tabularnewline
14 & $-$ & +  & $-$ & + &   & $-$ & $-$ & + & $-$ & (1,111) \tabularnewline
16 & + & $-$  & + & $-$ &   & $-$ & + & + & + & (1,000) \tabularnewline
\hline 
\hline 
\end{tabular}}
\label{Z2}
\end{table}
%%%

\newpage

\begin{figure}[htbp]
\includegraphics[width=86mm]{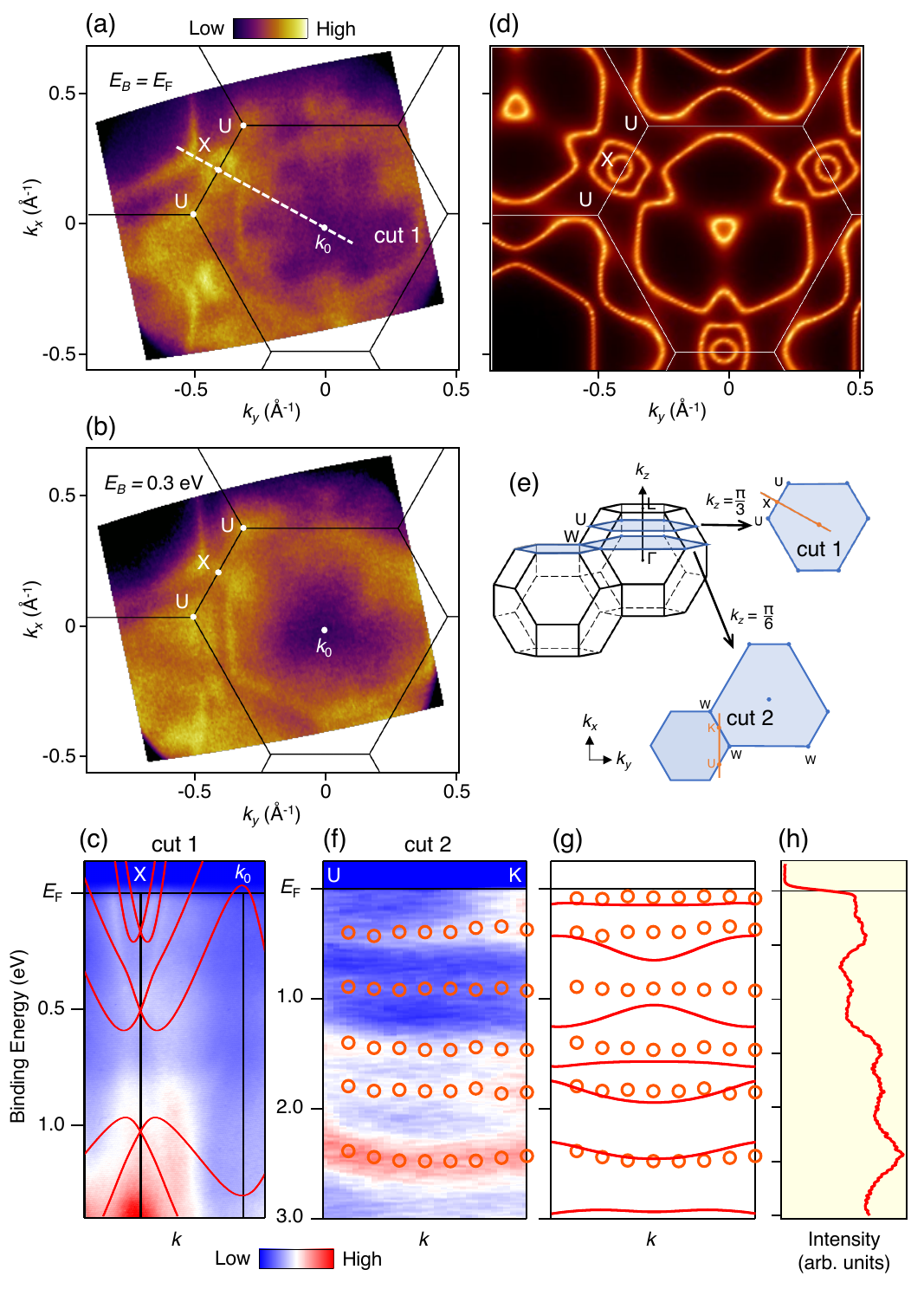}
 \caption{\label{Fig2}
(a) ARPES-intensity map at $E_F$ as a function of $\textit{k}_{x}$ and $\textit{k}_{y}$ at $\textit{k}_{z}$ $\sim$ $\pi$/3, measured with 21-eV photons. (b) Same as (a), but at $E_{\textit{B}}$ = 0.3 eV. (c) ARPES intensity and calculated band structure along cut 1 [white dashed line in (a)]. (d) Calculated Fermi surface at $\textit{k}_{z}$ = $\pi$/3. (e) Schematic of $\textit{k}_{z}$ = $\pi$/3 and $\pi$/6 planes in the bulk BZ, displayed with (left) 3D and (right) top views. (f) ARPES intensity along U-K [cut 2 in (e)], measured with 120-eV photons. (g) Calculated band structures along U-K. Orange circles in (f) and (g) show peak positions of EDCs. (h) Integrated EDC obtained with data in (f).}
\end{figure} 

\newpage

\begin{figure*}[htbp]
\includegraphics[width=165mm]{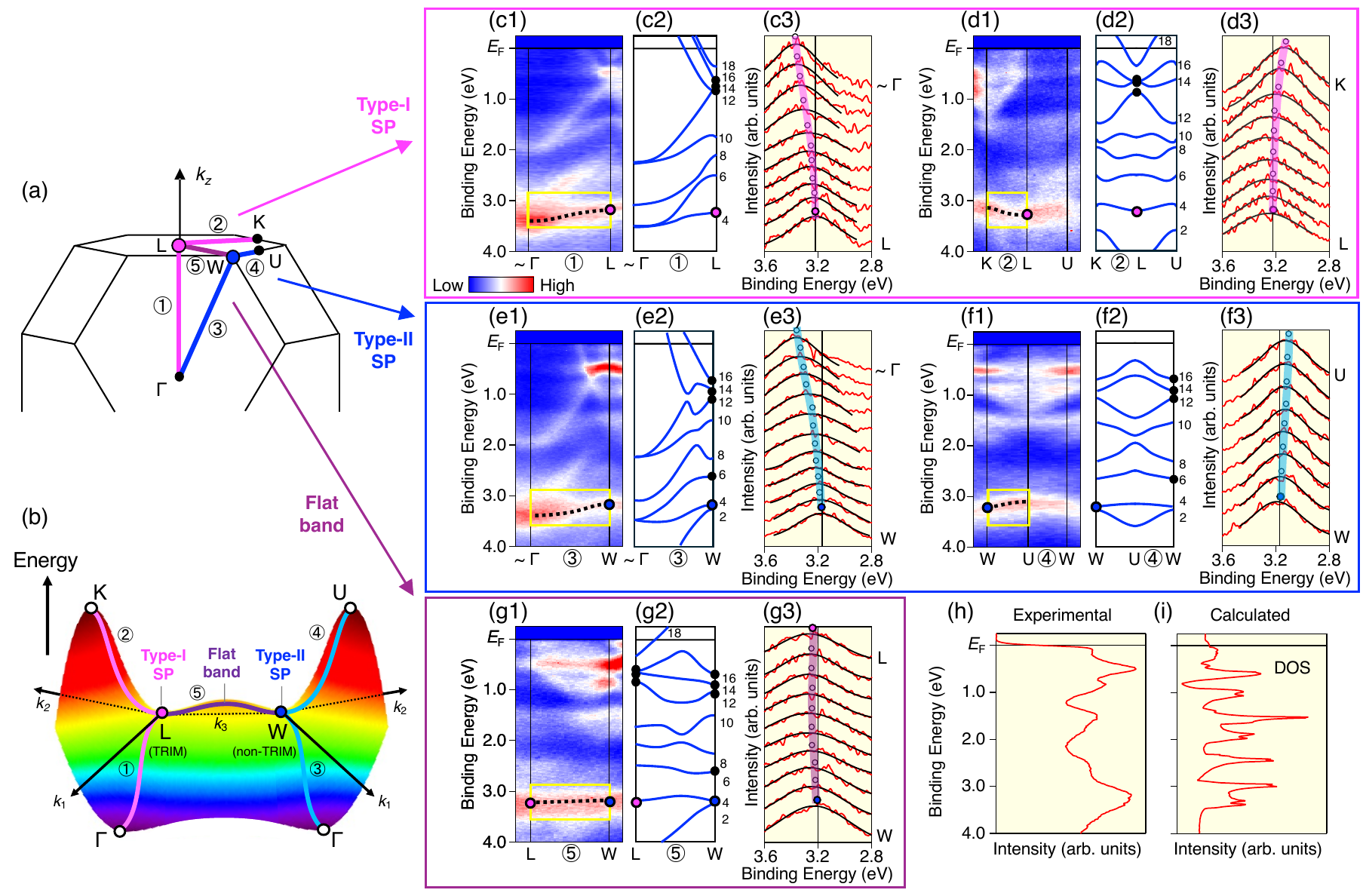}
 \caption{\label{Fig3}
(a) BZ of CsBi$_2$. (b) Schematic of type-I and type-II SPs at the L and W points, respectively, connected by a flat dispersion for band 4. (c) Plots of (c1) ARPES intensity along $\sim\!\Gamma$-L, (c2) corresponding calculated band structure, and (c3) EDCs in the yellow rectangle region in (c1). Black dashed line in (c1) is a guide for the eyes to trace band 4. Circles in (c3) represent peak positions of EDCs. (d)--(g) Same as (c), but along L-K/U, $\sim\!\Gamma$-W, W-U, and L-W, respectively. (h) Integrated EDC of (g1). (i) Calculated total DOS of CsBi$_2$ [same as the black curve in Fig. 1(f)].
}
\end{figure*}

\end{document}